# Spin current relaxation time in thermally evaporated pentacene films

Yasuo Tani,[1] Takuya Kondo,[1] Yoshio Teki,[2] and Eiji Shikoh[1,a)]

[1]*Graduate School of Engineering, Osaka City University, 3-3-138 Sugimoto, Sumiyoshi-ku, Osaka 558-8585, Japan*

[2]*Graduate School of Science, Osaka City University, 3-3-138 Sugimoto, Sumiyoshi-ku, Osaka 558-8585, Japan*

The spin current relaxation time ($\tau$) in thermally evaporated pentacene films was evaluated with the spin-pump-induced spin transport properties and the charge current transport properties in pentacene films. Under an assumption of a diffusive transport of the spin current in pentacene films, the zero-field mobility and the diffusion constant of holes in pentacene films were experimentally obtained to be $\sim 8.0 \times 10^{-7}$ m$^2$/Vs and $\sim 2.0 \times 10^{-8}$ m$^2$/s, respectively. Using those values and the previously obtained spin diffusion length in pentacene films of 42±10 nm, the $\tau$ in pentacene films was estimated to be 150±120 ns at room temperature. This estimated $\tau$ in pentacene films is long enough for practical use as a spintronic material.

[a)]E-mail: shikoh@elec.eng.osaka-cu.ac.jp

In spintronics,[1-2] to evaluate the spin transport properties in various materials are one of issues for the practical use. Organic molecules are promising materials for the spin transport because their spin-orbit interaction working as spin scattering centers is generally weak.[3] Until now, the spin transport properties of some molecular materials have been investigated.[4-9] We have focused on pentacene molecular films prepared by thermal evaporation as one candidate material for the spin transport, in terms of the good crystallinity even in the thermally evaporated films, and the high electrical conductivity without any dopants.[10] Moreover, pentacene shows the photo-conductivity for visible light,[11] where the spin transport properties of pentacene can be controlled through visible light irradiation. Recently, the spin transport in thermally evaporated pentacene films was achieved at room temperature (RT) with a pure spin current induced by spin-pumping, and the spin diffusion length ($\lambda$) in pentacene films was estimated to be about 42±10 nm.[9] The estimated $\lambda$ in pentacene films is comparable to that of other molecular materials evaluated with the method using the spin-pumping.[7,8] To precisely evaluate the spin transport properties of low conductive materials, the spin-pump-induced spin injection driven by ferromagnetic resonance (FMR) [12,13] is more effective than an electrical spin injection method. This is because the conductance mismatch problem at the interface between the ferromagnetic material as a spin injector and the target material for the spin injection is negligible in the spin injection by the spin-pumping,[7-9,14-19] whereas this problem lowers the spin injection efficiency in electrical spin injections.[20,21] In this study, by using the spin-pump-induced spin transport properties and the charge current transport properties in thermally evaporated pentacene films, the spin current relaxation time ($\tau$) in pentacene films was evaluated. When we say "spin relaxation time," in general, there are various spin relaxation times. In this study, we evaluated the spin relaxation time of the spin current which is the flow of the spin angular momentum. As other spin relaxation times, for example, there are the spin relaxation time for localized spins, the spin relaxation time that photo-excited spins are relaxed from the excited state to the ground state,[22] and so on. For the spin relaxation mechanism in the spin current, not only the spin-spin interaction and the spin-lattice interaction that are considered even in the spin relaxation of localized spins, but also the collision between the charges to carry the spin information must be considered (in this case, the charge is "hole"). Those effects can be taken into account via the experimentally obtained $\lambda$ of pentacene films.

In this study, the $\tau$ in the pentacene films is estimated as follows: because a diffusive transport of the spin current in the pentacene films has been assumed,[9] we can use a relationship of $\tau = \lambda^2/D$ [7,14] in thermally evaporated pentacene films, where the $D$ is the diffusion constant of holes in the pentacene films. The $\lambda$ is obtained from the spin transport experiments in pentacene films with changing the spin transport distance in the pentacene films. Our sample structure and experimental set up for the spin transport experiments in pentacene films is illustrated in Figure

1. An external static magnetic field ($H$) was applied with an angle ($\theta$) to the sample film plane. Spin transport in pentacene films is observed as follows: in a palladium(Pd)/evaporated-pentacene/Ni$_{80}$Fe$_{20}$ tri-layer sample, a pure spin current ($\vec{J}_S$), induced by the spin-pumping driven by the FMR of the Ni$_{80}$Fe$_{20}$ film is generated in the pentacene layer. This $\vec{J}_S$ is then absorbed into the Pd layer. The absorbed $\vec{J}_S$ is converted into a charge current as a result of the inverse spin-Hall effect (ISHE) [18] in Pd and detected as an electromotive force ($\vec{E}$),[14-19] which is expressed as,

$$\vec{E} \propto \theta_{SHE} \vec{J}_S \times \vec{\sigma}, \tag{1}$$

where $\theta_{SHE}$ is the spin-Hall angle, which is the efficiency of conversion from a spin current to a charge current, and $\vec{\sigma}$ is the spin-polarization vector in the $\vec{J}_S$. Thus, to detect an electromotive force due to the ISHE in Pd under the FMR of Ni$_{80}$Fe$_{20}$ is obvious evidence for spin transport in a pentacene film. Previously, we estimated the $\lambda$ in evaporated pentacene films of 42±10 nm with the evaluation of the pentacene layer thickness dependence of the spin transport.[9] Thus, in this letter, we only show a typical evidence of the spin transport in thermally evaporated pentacene films.

The Einstein relationship $D/\mu = k_B T/q$ is satisfied even in disordered organic molecular films,[23] where $\mu$, $k_B$, $T$ and $q$ are the zero-field charge mobility (not a field effect mobility) in pentacene films, the Boltzmann constant, the environmental temperature and the elemental charge (in the case of pentacene films, hole), respectively. The $\mu$ is obtained from measurements of the charge-current-density ($J$) versus the bias voltage ($V$) property in pentacene films. In an organic molecular film, the $J$ at low bias is approximately proportional to $V^2$, which is called the space charge limited current (SCLC) as follows:[24]

$$J = \frac{9}{8} \varepsilon \mu \frac{V^2}{L^3}, \qquad (2)$$

where $\varepsilon$ is the electric permittivity in the organic molecular film. The $L$ is charge transport distance in the organic molecular film. In short, to estimate the $\tau$ in pentacene films in this study, the $\mu$ in pentacene films is needed to measure, while the $\lambda$ has already been obtained.[9]

The sample preparation method for the spin transport experiments is as follows: Electron beam (EB) deposition was used to deposit Pd (Furuuchi Chemical Co., Ltd., 99.99% purity) to a thickness of 10 nm on a thermally oxidized silicon substrate, under a vacuum pressure of <$10^{-6}$ Pa. Next, also under a vacuum pressure of <$10^{-6}$ Pa, pentacene molecules (Sigma Aldrich Co., Ltd.; sublimed grade, 99.9%) were thermally evaporated to a thickness of 50 nm through a shadow mask. The pentacene deposition rate was varied from 0.03 nm/s to 0.40 nm/s. In this deposition rate range, however, no clear difference on our experiments was observed. Therefore, in this paper, we show the results with the pentacene deposition rate of 0.10 nm/s which is the same as that of our previous study.[9] The substrate temperature during pentacene depositions was RT. That deposition rate and the substrate temperature are similar conditions under which pentacene films show high crystallinity.[25] Finally, $Ni_{80}Fe_{20}$ (Kojundo Chemical Lab. Co., Ltd., 99.99%) was deposited to a thickness of 25 nm by EB deposition through another shadow mask, under a vacuum pressure of <$10^{-6}$ Pa. During $Ni_{80}Fe_{20}$ deposition, the sample substrate was cooled with a cooling medium of -2°C, to prevent the deposited molecular films from breaking.

To observe the *J-V* property in pentacene films, to use the same sample as that for the spin transport experiments is the best. However, to do it, one of leading wires to apply the electrical voltage to a sample must be attached on the $Ni_{80}Fe_{20}$ film with the thickness of only 25 nm formed on Pd/pentacene films. In general, that is difficult because the very thin $Ni_{80}Fe_{20}$ film and the under-layers are broken at the wire bonding procedure. Therefore, we prepared another type sample to measure the electrical properties in pentacene films as follows: pentacene molecules were thermally evaporated on a couple of pre-patterned $Ni_{80}Fe_{20}$ electrodes on a thermally oxidized silicon substrate with a conventional photolithography technique and EB deposition. The gap length between the two electrodes, which corresponds to the *L* in pentacene films, was designed to be 2 μm. The evaporation condition of the pentacene molecules was the same as that of samples for the spin transport experiments. The cross sectional area of the charge transport of pentacene films is 30 nm (the electrode thickness) × 3 mm (the pentacene layer width) = $9 \times 10^{-11}$ $m^2$.

For evaluation of the spin-pump-induced spin transport property in pentacene films, a microwave $TE_{011}$-mode cavity in a conventional electron spin resonance (ESR) system (JEOL, JES-TE300) to excite FMR in $Ni_{80}Fe_{20}$, and a nano-voltmeter (Keithley Instruments, 2182A) to detect electromotive forces from the samples were used. The ESR system equips with a goniometer to measure the $\theta$ dependent properties. Leading wires for detecting the output voltage properties were directly attached with silver paste at both ends of the Pd layer. For measurement of the *J-V* properties of pentacene films, a two-probe method was used with a DC voltage current source/monitor (ADVANTEST Corporation, R6243). All of the measurements were performed at RT.

Figure 2(a) shows the FMR spectrum of a sample at the $\theta$ of 0°, under a microwave power of 200 mW. The FMR field ($H_{FMR}$) of the $Ni_{80}Fe_{20}$ film is 1200 Oe at a microwave frequency (*f*) of 9.45 GHz, and the $4\pi M_s$ of the $Ni_{80}Fe_{20}$, where $M_s$ is the saturation magnetization of the $Ni_{80}Fe_{20}$ film, is estimated to be 7300 G under the FMR condition in the in-plane field:

$$\frac{\omega}{\gamma} = \sqrt{H_{FMR}(H_{FMR} + 4\pi M_S)}, \tag{3}$$

where $\omega$ and $\gamma$ are the angular frequency ($2\pi f$) and the gyromagnetic ratio of $1.86 \times 10^7$ $Oe^{-1}s^{-1}$ of $Ni_{80}Fe_{20}$, respectively.[9,19] Fig. 2(b) shows the output voltage properties of the same sample shown in Fig. 2(a); the circles represent experimental data and the solid lines are the curve fit obtained using the equation:[14,16-19]

$$V(H) = V_{ISHE} \frac{\Gamma^2}{(H-H_{FMR})^2 + \Gamma^2} + V_{Asym} \frac{-2\Gamma(H-H_{FMR})}{(H-H_{FMR})^2 + \Gamma^2}, \qquad (4)$$

where $\Gamma$ denotes the damping constant (110 Oe in this study). The first and second terms in eq. (4) correspond to the symmetry term to $H_{FMR}$ corresponding to the ISHE and the asymmetry term to $H_{FMR}$ (e.g. anomalous Hall effect [14,16-19] and other effects showing the same asymmetric voltage behavior relative to the $H_{FMR}$, like parasitic capacitances), respectively. $V_{ISHE}$ and $V_{Asym}$ correspond to the coefficients of the first and second terms in eq. (4). Output voltages are observed at $H_{FMR}$ at $\theta$ of 0 and 180°. Notably, the output voltage changes sign between $\theta$ values of 0 and 180°. This sign inversion of the output voltages in Pd associated with the magnetization reversal in $Ni_{80}Fe_{20}$ is characteristic of the ISHE.[14,16-19] The experimental results in this study are similar to the previous report,[9] that is, the spin-pump-induced spin transport in thermally evaporated pentacene films has experimentally been demonstrated. The detailed $\theta$ dependence of the $V_{ISHE}$ of the Pd in a sample is shown in the supplemental material.

Figure 3 shows (a) the $J$-$V$ property and (b) the $J$-$V^2$ property of a pentacene film. The dashed line in Fig. 3(b) shows an example of a linear fitting to the $J$-$V^2$ properties at lower bias, where a little deviation for the linear fittings exists. Using the relative electric permittivity in the pentacene films ($\varepsilon_r$) of ~3,[26] and eq. (2), the $\mu$ in pentacene films was estimated to be $0.40\sim1.2\times10^{-6}$ m$^2$/Vs at RT. Here, just for simplification, the difference between the work function of the $Ni_{80}Fe_{20}$ electrode and the ionization potential of the pentacene film is ignored because only the divergence of $J$-$V^2$ properties at low bias is needed for the analysis. The deviation on the $\mu$ calculations comes from the deviation of the divergence of the linear fitting to $J$-$V^2$ properties. Under an assumption of a diffusive transport of the spin current in the pentacene films, the $D$ in the pentacene films was estimated to be $1.0 \sim 3.0 \times 10^{-8}$ m$^2$/s with the Einstein relationship. Thus, using the relationship of $\tau = \lambda^2/D$ and the $\lambda$ of 42±10 nm,[9] the $\tau$ in the pentacene films at room temperature was estimated to be 150±120 ns. In the deviation of this estimated $\tau$, the deviation of the previously estimated $\lambda$ in pentacene films is also included.

We also estimated the spin relaxation time in the Hanle effect type spin precession in spin transport (See the supplemental material). The spin relaxation time in the Hanle effect type spin precession (hereafter, called as "Hanle-type spin relaxation time") in spin transport is a different physical phenomenon from that of the $\tau$. The Hanle-type spin relaxation time is the relaxation time that the spins propagating in a material orient from the initial direction to the applied magnetic field direction in the spin transport, and just after the relaxation, the spin information will be still alive. Meanwhile, the spin current relaxation time ($\tau$) is defined as the relaxation time that the spin signal reflecting the spin propagation in a material decays to 1/$e$ in the spin transport, and just after the relaxation, the spin information is almost dissipated. The values of the two spin relaxation times related to the spin transport may be different and may be coincident, depending on materials. If the Hanle-type spin relaxation time is estimated to be below ~1 ns, the estimation of the Hanle-type spin relaxation time is reliable (see the supplementary material). However, if the Hanle-type spin relaxation time is over 1 ns, the analysis to estimate the Hanle-type spin relaxation time may not be reliable in the method [7] because the fitting curves for the estimation of the Hanle-type spin relaxation time do not change (see the supplementary material). In this study, the Hanle-type spin relaxation time was estimated to be >1ns. Thus, we didn't know whether the spin current relaxation time ($\tau$) and the Hanle-type spin relaxation time were coincident or not.

In the last part of this manuscript, we check the validity of the estimated $\tau$ in the pentacene films from the two viewpoints. One is the comparison with the spin relaxation time that photo-excited spins are relaxed from the excited state to the ground state, over several tens μs.[22] This is much longer time than the estimated $\tau$ of 150±120 ns. This difference might come from that the collision between charges in the relaxation process is considered or not. Therefore, the estimated $\tau$ in pentacene films of 150±120 ns is valid. Next, we compared the estimated $\tau$ with the $\tau$ values in other molecular materials evaluated with the spin-pump-induced spin transport experiments.[7,8] The $\tau$ for a conductive polymer PBTTT has been reported to be ~20 ms,[7] which is extremely long. However, the spin propagation in the PBTTT is thought to be due to the transport by polarons, which is a different spin transport mechanism from that in pentacene films. The $\tau$ for a $p$-type polymeric semiconductor PEDOT:PSS has been estimated to be ~30 ns,[8] which is roughly comparable to the result of this study. The estimated $\tau$ in pentacene films of 150±120 ns is long enough for practical use as a spintronic material, comparing with the $\tau$ of n-type silicon of ~1.3 ns [27] and the $\tau$ of p-type silicon of ~120 ps.[14] For example, this $\tau$ of 150±120 ns is long enough to control the spin transport property in pentacene films through visible light irradiation.

In summary, we evaluated the $\tau$ in thermally evaporated pentacene films with the spin-pump-induced spin transport properties and the charge current transport properties in pentacene films. Using the $J$-$V$ property in pentacene films, the $\mu$ in pentacene films was estimated to be $0.40 \sim 1.2 \times 10^{-6}$ m$^2$/Vs. Under an assumption of a diffusive transport of the spin current in pentacene films, the $D$ of pentacene films, was obtained to be $1.0 \sim 3.0 \times 10^{-8}$ m$^2$/s. By using the relationship of $\tau = \lambda^2/D$, the $\tau$ in pentacene films at room temperature was estimated to be 150±120 ns. This estimated $\tau$ in pentacene films is long enough for practical use as a spintronic material.

SUPPLEMENTARY MATERIAL

See supplementary material for the $\theta$ dependence of the $V_{\text{ISHE}}$ of the Pd in a sample and the estimation of the spin relaxation time in the Hanle effect type spin precession in spin transport.

ACKNOWLEDGEMENT

This research was partly supported by a Grant-in-Aid from the Japan Society for the Promotion of Science (JSPS) for Scientific Research (B) (26286039).

References

[1] R.L. Stamps, S. Breitkreutz, J. Åkerman, A.V. Chumak, Y. Otani, G.E.W. Bauer, J.-U. Thiele, M. Bowen, S.A. Majetich, M. Kläui, I.L. Prejbeanu, B. Dieny, N.M. Dempsey, and B. Hillebrands, J. Phys. D: Appl. Phys. **47**, 333001 (2014).

[2] A. Hoffmann and S.D. Bader, Phys. Rev. Appl. **4**, 047001 (2015)

[3] M. Shiraishi and T. Ikoma, Physica E **43**, 1295 (2011).

[4] Z.H. Xiong, Di Wu, Z.V. Vardeny, and J. Shi, Nature **427**, 821 (2004).

[5] X. Sun, M. Gobbi, A. Bedoya-Pinto, O. Txoperena, F. Golmar, R. Llopis, A. Chuvilin, F. Casanova, and L. Hueso, Nature Commun. **4**, 2794 (2013).

[6] T. Ikegami, I. Kawayama, M. Tonouchi, S. Nakao, Y. Yamashita, and H. Tada, Appl. Phys. Lett. **92**, 153304 (2008).

[7] S. Watanabe, K. Ando, K. Kang, S. Mooser, Y. Vaynzof, H. Kurebayashi, E. Saitoh, and H. Sirringhaus, Nature Phys. **10**, 308 (2014).

[8] M. Kimata, D. Nozaki, Y. Niimi, H. Tajima, and Y. Otani, Phys. Rev. B **91**, 224422 (2015).

[9] Y. Tani, Y. Teki, and E. Shikoh, Appl. Phys. Lett. **107**, 242406 (2015).

[10] H. Cheng, Y. Mai, W. Chou, and L. Chang, Appl. Phys. Lett. **90**, 171926 (2007).

[11] O. Ostroverkhova, D.G. Cooke, S. Shcherbyna, R.F. Egerton, F.A. Hegmann, R.R. Tykwinski, and J.E. Anthony, Phys. Rev. B **71**, 035204 (2005).

[12]S. Mizukami, Y. Ando, and T. Miyazaki, Phys. Rev. B **66**, 104413 (2002).

[13]Y. Tserkovnyak, A. Brataas, and G.E.W. Bauer, Phys. Rev. Lett. **88**, 117601 (2001).

[14]E. Shikoh, K. Ando, K. Kubo, E. Saitoh, T. Shinjo, and M. Shiraishi, Phys. Rev. Lett. **110**, 127201 (2013).

[15]R. Ohshima, A. Sakai, Y. Ando, T. Shinjo, K. Kawahara, H. Ago, and M. Shiraishi, Appl. Phys. Lett. **105**, 162410 (2014).

[16]A. Yamamoto, Y. Ando, T. Shinjo, T. Uemura, and M. Shiraishi, Phys. Rev. B **91**, 024417 (2015).

[17]S. Dushenko, M. Koike, Y. Ando, T. Shinjo, M. Myronov, and M. Shiraishi, Phys. Rev. Lett. **114**, 196602 (2015)

[18]E. Saitoh, M. Ueda, H. Miyajima, and G. Tatara, Appl. Phys. Lett. **88**, 182509 (2006).

[19]K. Ando and E. Saitoh, J. Appl. Phys. **108**, 113925 (2010).

[20]G. Schmidt, D. Ferrand, L.W. Molenkamp, A.T. Filip, and B.J. van Wees, Phys. Rev. B **62**, R4790 (2000).

[21]A. Fert and H. Jaffres, Phys. Rev. B, **64**, 184420 (2001).

[22]T. Kawahara, S. Sakaguchi, K. Tateishi, T.L. Tang, and T. Uesaka, J. Phys. Soc. Jpn. **84**, 044005 (2015).

[23]G.A.H. Wetzelaer, L.J.A. Koster, and P.W.M. Blom, Phys. Rev. Lett. **107**, 066605 (2011).

[24] P.W.M. Blom, M.J.M. de Jong, and M.G. van Munster, Phys. Rev. B. **55**, R656 (1997).

[25] C.D. Dimitrakopoulos and P.R.L. Malenfant, Adv. Mater. **14**, 99 (2002).

[26] C.H. Kim, O. Yaghmazadeh, D. Tondelier, Y.B. Jeong, Y. van Bonnassieux, and G. Horowitz, J Appl. Phys. **109**, 083710 (2011).

[27] T. Suzuki, T. Sasaki, T. Oikawa, M. Shiraishi, Y. Suzuki, and K. Noguchi, Appl. Phys. Exp. **4**, 023003 (2011).

Figure captions:

FIG. 1. (Color online) Schematic illustrations of our sample and orientation of external applied magnetic field ($H$) used in the experiments. $J_S$ and $E$ correspond, respectively, to the spin current generated in the pentacene film by the spin-pumping and the electromotive forces due to the ISHE in Pd. $\theta$ is the $H$ angle to the sample film plane.

FIG. 2. (Color online) (a) FMR spectrum and (b) output voltage properties of a typical sample under a microwave power of 200 mW.

FIG. 3. (Color online) Current-density ($J$) versus bias-voltage ($V$) property of a pentacene film, plotted by (a) $J$-$V$ and (b) $J$-$V^2$. At low bias, the $J$ is proportional to $V^2$ and the behavior is drawn as a dashed line in (b).

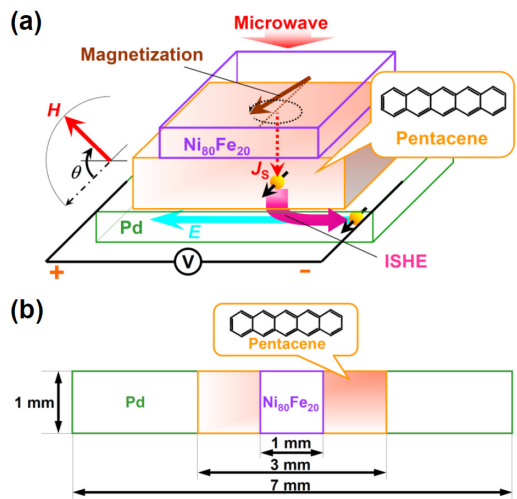

Y. Tani, et al.:  FIG. 1.

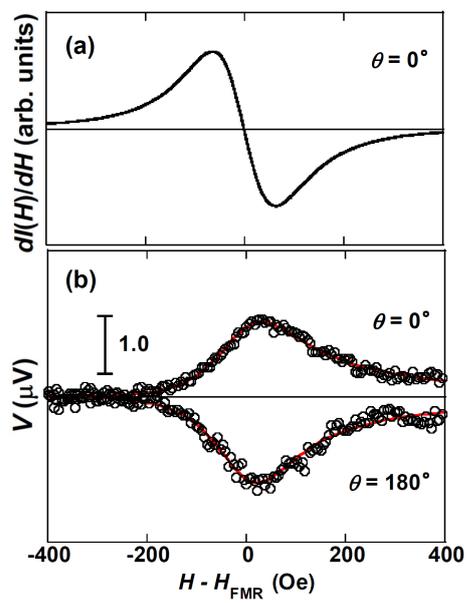

Y. Tani, et al.: FIG. 2.

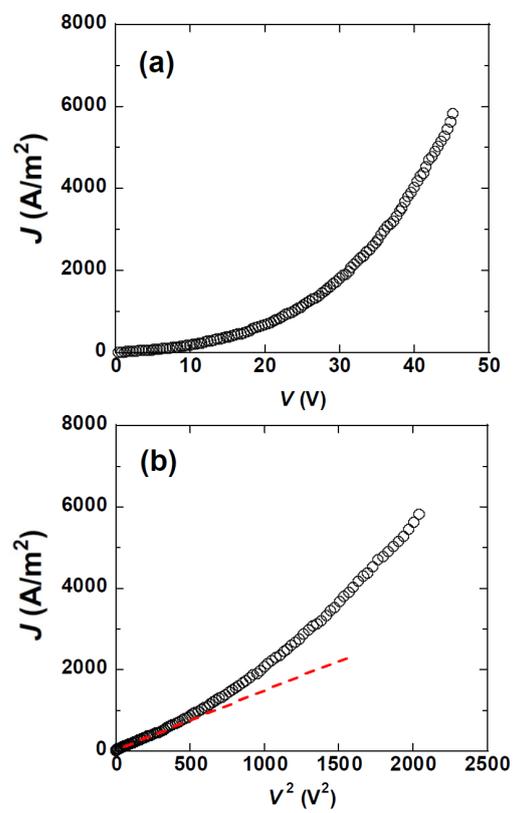

Y. Tani, et al.: FIG. 3.